\documentclass[useAMS,usenatbib,usegraphicx]{mn2e}

\newcommand{\nc}{\newcommand}

\nc{\be}[1]{\begin{equation}\mbox{$\label{#1}$}}
\nc{\bea}[1]{\begin{eqnarray} \mbox{$\label{#1}$}}
\nc{\Section}[2]{\section{#2}\label{#1}}
\nc{\Bibitem}[1]{\bibitem{#1}}
\nc{\Label}[1]{\label{#1}}

\nc{\Mpc}{Mpc/h}
\nc{\vev}[1]{\langle #1 \rangle}

\nc{\eea}{\end{eqnarray}}
\nc{\ee}{\end{equation}}

\def\lcdm{$\Lambda$CDM~}

\def\etal{{et al. }}
\def\etals{{et al. }}
\def\ltsima{$\; \buildrel < \over \sim \;$}
\def\gtsima{$\; \buildrel > \over \sim \;$}
\def\simlt{\lower.5ex\hbox{\ltsima}}
\def\simgt{\lower.5ex\hbox{\gtsima}}
\nc{\w}{$w_2(\theta)$\ }
\nc{\ie}{i.e.}
\nc{\eg}{e.g.}
\def\q{{\bf q} }

\title[Quadrupole in WMAP]
{2-point anisotropies in WMAP and the Cosmic Quadrupole}

\author[Gazta\~naga, Wagg, Multam\"aki, Monta\~na, Hughes]
{E.Gazta\~naga$^{1,2}$.
J.Wagg$^2$, T.Multam\"aki$^3$, 
A.Monta\~na$^2$,  D.H.Hughes$^2$\\
\\
$^1$Institut d'Estudis Espacials de Catalunya, IEEC/CSIC, Gran
Capit\'an 2-4, 08034 Barcelona, Spain\\
$^2$Instituto Nacional de Astrof\'{\i}sica, \'Optica y Electr\'onica
(INAOE), Aptdo. Postal 51 y 216, Puebla, Mexico \\
$^3$Departament E.C.M. and C.E.R. 
, Universitat de Barcelona, Diagonal 647, 08028 Barcelona, Spain\\
}

\date{Version 14 April 2003}

\pagerange{\pageref{firstpage}--\pageref{lastpage}} 
\pubyear{2003}

\begin{document}

\label{firstpage}

\maketitle

\begin{abstract}
Large-scale modes in the temperature anisotropy power spectrum $C_l$
measured by the Wilkinson Microwave Anisotropy Probe (WMAP), seem to
have lower amplitudes ($C_2$, $C_3$ and $C_4$) than that expected in
the so called concordance \lcdm model. In particular, the quadrupole
$C_2$ is reported to have a smaller value than allowed by cosmic
variance.  This has been interpreted as a possible indication of new
physics. In this paper we re-analyse the WMAP data using the 2-point
angular correlation and its higher-order moments. This method, which
requires a full covariance analysis, is more direct and provides
better sampling of the largest modes than the standard harmonic
decomposition.  We show that the WMAP data is in good agreement
($\simeq 30\%$ probability) with a \lcdm model when the WMAP data is
considered as a particular realization drawn from a set of realistic
\lcdm simulations with the corresponding covariance.  This is also
true for the higher-order moments, shown here up to 6th order, which
are consistent with the Gaussian hypothesis. The sky mask plays
a major role in assessing the significance of these agreements.
We recover the best
fit model for the low-order multipoles based on the 2-point
correlation with different assumptions for the covariance.  Assuming
that the observations are a fair sample of the true model, we find
$C_2 = 123 \pm 233$, $C_3= 217 \pm 241 $ and $C_4 = 212 \pm 162$ (in
$\mu K^2$).  The errors increase by about a factor of 5 if
we assume the \lcdm model.  If we exclude the Galactic plane $|b|<30$
from our analysis, we recover very similar values within the errors
(ie $C_2=172, C_3= 89, C_4=129 $). This indicates that the Galactic
plane is not responsible for the lack of large-scale power in the WMAP
data.

\end{abstract}

\begin{keywords}
cosmology -- cosmic microwave background
\end{keywords}


\section{Introduction}

Measurements of the cosmic microwave background (CMB)
anisotropies by  WMAP (Bennett et al. 2003; Spergel \etal 2003; 
Peiris \etals 2003) are in good agreement with a `concordance' 
cosmology based on the \lcdm model.
However, the WMAP results also confirm the low amplitude of the CMB
quadrupole first measured by COBE (eg Hinshaw \etals 1996a). As Bennett \etals
(2003) comment, the amplitudes of the quadrupole and 
the octopole are low compared with the predictions of \lcdm models.
The WMAP team also present a convincing
case that the low CMB multipoles are not significantly affected by
foreground Galactic emission (see also Tegmark, de Oliveira-Costa and
Hamilton 2003). The discrepancy between the observations and the
\lcdm model is particularly evident 
in the temperature angular correlation function, which shows an
almost complete lack of signal on angular scales $\simgt 60$ degrees.
According to Spergel \etals (2003), the probability of finding such a
result in a spatially-flat \lcdm cosmology is about $1.5 \times
10^{-3}$. This is small enough to require an explanation,
e.g. in the form of new physics \cite{conta}
or in terms of spatial curvature \cite{efs03}. 

In this paper we consider the WMAP data from the point of view of the
two-point angular correlation function as opposed to the usual
harmonic decomposition. In some aspects our approach is similar to the
study presented by Hinshaw \etal (1996b) for the COBE data.  The paper
is organized as follows.  In \S 2 we introduce our methodology
and test it with realistic simulations. In \S 3 we apply our method to
the real WMAP data and present a comparison with the simulations. We
also describe the application of higher-order statistics to test
the Gaussianity of WMAP. The conclusions are described in \S 4.


\section{2-pt correlation $w_2(\theta)$ in \lcdm models}

On the largest scales, the 2-point correlation $w_2(\theta)$ has some
important advantages and disadvantages over spherical harmonics.  The
overall shape of the 2-point function is very sensitive to small
differences in power at low multipoles. This is clearly an advantage
when measuring the quadrupole. On the other hand, different bins in
$\theta$ are highly correlated, which means that we need to use the
full covariance matrix to assess the significance of any departures
between measurements and models.  Another important advantage of using
the 2-point function is that large-scale modes can be easily sampled
from any region of the sky, even when a mask is required, provided the
region is large enough. In comparison, harmonic decomposition breaks
the angular symmetry in the sky into orthogonal bases which have an
arbitrary phase orientation. This is not a problem in the case of full
sky coverage, but the results will be coordinate-dependent if a mask
is used.  Moreover, the harmonic coefficients $a_{lm}$ (see
Eq.\ref{harmdec}) over masked data are convolved with the harmonic
image of the mask. The resulting coupling matrix needs to be inverted
and the resulting estimator for the recovered coefficients is
typically biased because of the non-linear transformations
involved. These potential complications are avoided in configuration
space.

\subsection{Definition and estimators}

The 2-point angular correlation function is defined as the 
expectation value or mean cross-correlation of density fluctuations 
at two positions $\q_1$ and $\q_2$ in the sky:
\be{w2def}
w_2(\theta) \equiv \vev{ \Delta T({\bf\q_1}) \Delta T({\bf\q_2}) },
\ee
where $\theta = |\bf{\q_2}-\bf{\q_1}|$, assuming that
the distribution is statistically isotropic. 

To estimate $w_2(\theta)$ from the pixel maps we use a simple
but effective estimator:
\be{w2est}
w_2(\theta) = {\sum_{i,j} \Delta_i \Delta_j ~w_i~w_j \over{\sum_{i,j} w_i~w_j}},
\ee
where $\Delta_i$ and $\Delta_j$ are the temperature differences in the map
and the sum extends to all pairs separated by $\theta \pm \Delta\theta$. 
The differences are normalised so that $\vev{\Delta_i}=0$.
The weights $w_i$ can be used to minimise the variance when the pixel
noise is not uniform, but this introduces larger cosmic variance.
Here we follow the WMAP team and use uniform weights (i.e. $w_i=1$). 
We use bins $\Delta\theta$ whose size is proportional to the square root of 
the angle $\theta$.

This estimator is unbiased and is equivalent to the minimum variance
estimator of Landy \& Szalay (1993). For higher-order moments see \S 3.6.

\subsection{The \lcdm Model and simulations}

Temperature fluctuations can be expanded in terms of spherical
harmonics according to,

\be{harmdec}
\Delta T(\q) = \sum_{\ell=0}^{\infty}\sum_{m=-\ell}^{\ell} a_{\ell m} Y_{\ell}^{m}(\q),
\ee
which defines the multipole power spectrum: $C_l = <|a_{\ell m}|^2>$.
Gaussian simulations can then be made by taking $a_{\ell m}$ to be independent
Gaussian fields with random phases, zero mean and variance $C_l$.

\begin{figure}
\vskip 3 truein
\includegraphics{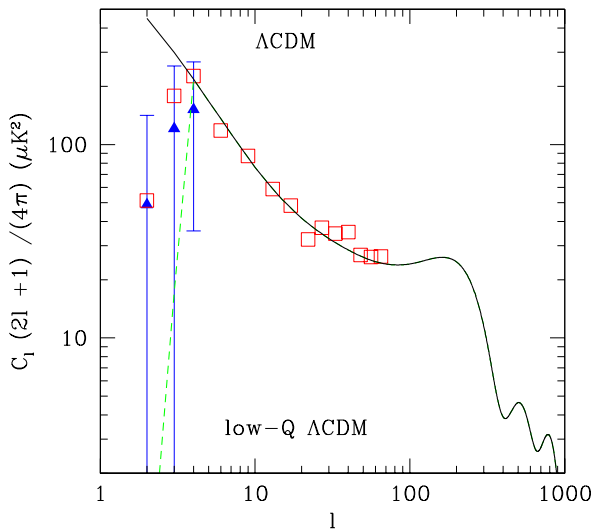}
\caption
{Power spectrum $C_l$ for the \lcdm model (continuous line).
Open squares show the first multipoles by Bennett et al. (2003).
Closed triangles with errors correspond to Eq.[\ref{eq:wmapcls}].
Unlike the conventional way of plotting this curve, each
amplitude here shows the contribution of each multipole to the sky 
anisotropies in Eq. [\ref{eq:w2cl}]. Dashed line shows
the low-Q model in Eq. [\ref{eq:low-Q}].}
\label{cl.eps}
\end{figure}

In order to test the significance of the correlations measured in the
WMAP data, we make simulated sky maps by using the best-fit
cosmological parameters estimated from WMAP (Spergel \textit{et al.}
2003). The theoretical power spectra are generated using the publicly
available CMBFAST package (Seljak \& Zaldarriaga 1996). The 
$C_l$s are normalised to the COBE spectra and the input parameters are listed in Table
\ref{tab:table1}. Along with cosmological parameters, CMBFAST accepts
a reionization optical depth, $\tau$, and the scalar spectral
index, $\eta_s$.  From now on we refer to this particular set of paramenters
 as the \lcdm model.
Figure \ref{cl.eps} shows the resulting power spectrum for the \lcdm model.
In the \lcdm model the first three non-zero multipoles are (in $\mu K^2$):
\be{eq:cls}
C_2=1130,\ \ C_3=537,\ \ C_4=304,
\ee
while Bennett et al. (2003) find from WMAP data:
\bea{eq:wmap2003cls}
C_2 & = & 129 \pm 799\nonumber\\
C_3 & = & 320 \pm 320\nonumber\\
C_4 & = & 316 \pm 104.
\eea

\begin{table}
\caption{CMBFast Cosmological Parameters
\label{tab:table1}}
\begin{center}
\begin{tabular}{cc} \hline \hline
\smallskip
Parameter & Value \\
H$_0$ & 72 km/s/Mpc \\
$\Omega_b$ & .046 \\
$\Omega_c$ &  .224 \\
$\Omega_\Lambda$ & .730 \\
$\Omega_k$ & 0.0 \\
$\tau$ & .17 \\
$\eta_s$ & 1.0 \\
\hline
\end{tabular}
\end{center}
\noindent
\end{table}

This power spectrum is then used as input to generate all-sky CMB maps
using the HEALPix\footnote{http://www.eso.org/science/healpix/}
package (G\'orski, Hivon \& Wandelt 1999). We simulate 100 V-band maps
(FWHM = 21' at 61\,GHz), each with a different random phase, using the
best fit WMAP power spectra\footnote{We have also generated 20 W-band
simulations to compare with W-band WMAP, but find no significance
differences with V-band and decided to use only the latter because of
the smaller pixel noise.}  These maps are all generated with the RING
pixelization scheme and nside=512.  Noise is added to the simulations
following the pixel-noise pattern in the WMAP data. We also impose the most
conservative WMAP kp0 foreground mask to the simulations.

For some crucial aspects of the analysis we have used 
an additional set of 2000
V-band \lcdm simulations to check how the results depend on the
number  of simulations used. 
In general we find that with 20 simulations one finds similar
qualitative conclusions to those with 2000 simulations,
with a typical uncertainty of
$\simeq 50\%$ in the intervals of confidence (see below \S3.4).


\subsection{The low-Q \lcdm model}

As a useful comparison model, we design a fiducial variation of the 
\lcdm model with low values of the quadrupole and octopole. This new
model is identical to the \lcdm model for $l>3$ but has zero quadrupole 
$C_2=0$ and half the octopole $C_3\rightarrow C_3/2$. As we will show later,
this model is in better agreement with the WMAP data than the \lcdm model
and follows the curved universe with a truncated spectrum \cite{efs03}.
Thus, we modify the \lcdm power spectrum by setting:
\bea{eq:low-Q}
C_{2}^{low-Q} & = & 0 \\
C_{3}^{low-Q} & = & \frac 12 C_3^{\Lambda CDM}\nonumber\\
C_{l}^{low-Q} & = & C_l^{\Lambda CDM},\ l>3\nonumber.
\eea
We refer to this model as {\it low-Q \lcdm}.
These new $C_{l}$'s are then used to generate 20 HEALPix V-band maps, 
each with a different random phases. 

\subsection{Predicted $w_2(\theta)$ from $C_l$}

The 2-pt function, $w_2(\theta)$, is the Fourier transform of the
angular power spectrum of temperature fluctuations. In terms of the 
$C_l$ in Eq. [\ref{harmdec}], 
we have (see e.g. Bond \& Efstathiou 1987):
\be{eq:w2cl}
w_2(\theta) = {1\over{4\pi}} \sum_{l=2}~ (2 l +1) ~C_l ~P_l(\cos{\theta}),
\ee
where $P_l$ are the Legendre polynomials. The sum starts at $l=2$ because 
the monopole and the dipole have been subtracted from the WMAP data 
(CMB rest frame). 

\begin{figure}
\vskip 3 truein
\includegraphics{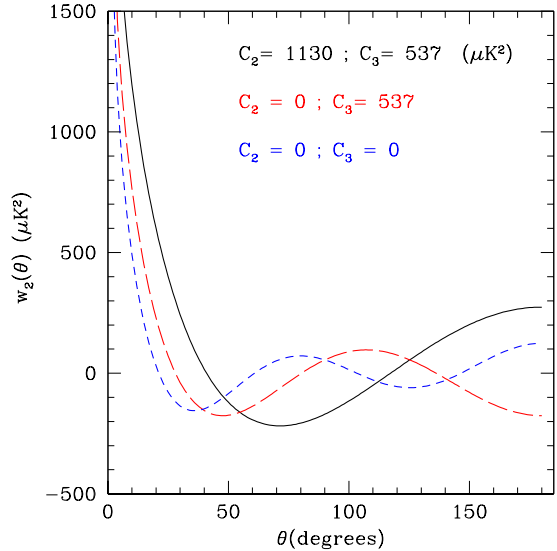}
\caption
{Theoretical prediction of the 2-point angular correlation $w_2$ in the \lcdm
model with all multipoles (continuous line), without the quadrupole
(long dashed line) and without quadrupole and octopole
(short dashed line).}
\label{w2low}
\end{figure}

Figure \ref{cl.eps} shows the multipole coefficients for the \lcdm
and low-Q models. Figure \ref{w2low} shows the corresponding analytical
summation for all multipoles $l \ge 2$ (continuous line), $l \ge 3$ (long-dashed line) and
$l \ge 4$ (short-dashed line). The figure very clearly illustrates the 
sensitivity of the angular 2-point function to the low multipoles:
e.g. varying just the quadrupole completely changes the shape of the curve
at large scales.

\subsection{$w_2(\theta)$ from simulations}

\begin{figure}
\vskip 3.3 truein
\includegraphics{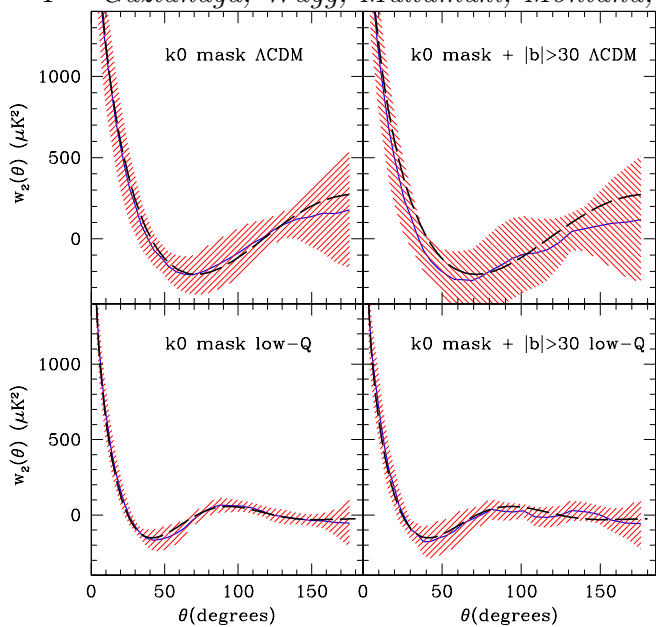}
\caption
{The 2-pt function $w_2(\theta)$ from simulations (continuous line)  with 1-sigma
confidence region (shaded region) compared to $w_2(\theta)$ from the input
$C_l$ spectrum (dashed line). Top left: \lcdm model with kp0-mask. Top right:
\lcdm model with kp0-mask and $|b|>30$ Galactic cut. Bottom left: 
low-Q model  with kp0-mask. Bottom right:
low-Q model  with kp0-mask and $|b|>30$ Galactic cut.}
\label{w2sim.eps}
\end{figure}

Figure \ref{w2sim.eps} compares the theoretical values of $w_2(\theta)$ in the 
\lcdm and low-Q models with that estimated from the simulated maps.
  The continuous line 
is the mean of the different realizations of each model and the dashed line
is the theoretical estimate corresponding to the set of $C_l$s. 
The shaded area shows the 1-sigma dispersion of the simulations in each bin. As can be seen
from the figure, the \lcdm model has the larger errors. This is expected
since the \lcdm model has stronger lower-order multipoles which in turn 
lead to larger variations from realization to realization.

The right-hand panels in Figure \ref{w2sim.eps} show the estimation 
of $w_2$ for pixels 
within Galactic latitudes $|b|>30$ degrees. This reduces 
the number of 
pixels by almost a factor of two which results in larger errors. Note how
the recovered values of $w_2$ are not strongly biased within the error bars. 
Thus, even after excluding a broad Galactic region one can still use
$w_2$ to separate a low and high quadrupole in the data. We will further 
elaborate on this point later in the paper.

\subsection{Error bars}

\begin{figure}
\vskip 3 truein
\includegraphics{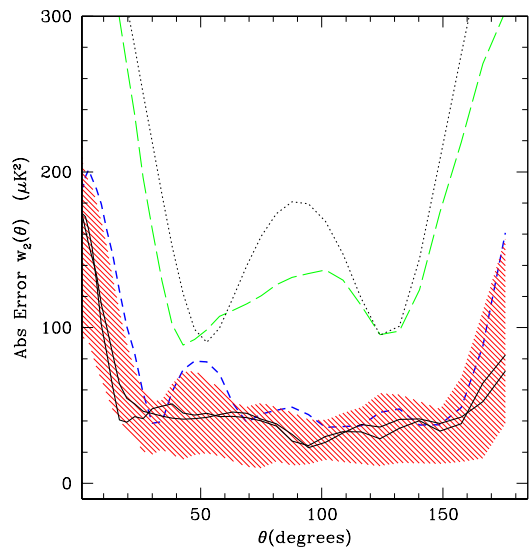}
\caption
{Short, long-dashed and dotted lines show the absolute errors (or dispersion) 
in $w_2$ from the low-Q simulations, 
the first 40 and the first 1000 \lcdm simulations respectively.
The continuous lines corresponds to estimated jackknife errors with N=8
and N=32 subsamples within the 
WMAP data, while the shaded region shows the  dispersion in jackknife errors from
each of the low-Q realizations.}
\label{ew2vmap}
\end{figure}

The absolute error (1-sigma dispersion) as a function of scale from
 different realizations of each model is shown in Figure
 \ref{ew2vmap}. As can be seen from the figure, errors for the \lcdm
 simulations (long-dashed lines) are much larger than errors in the
 low-Q model (short-dashed line). Note also how the low-Q model
 estimation is more noisy since it is based on 20 realizations, while
 there are 100 and 1000 realizations in the \lcdm model.

The shaded region represents
the dispersion (2-sigma) in the jackknife errors (see below) for the
different low-Q realizations. 

\subsection{Covariance Matrix}

As mentioned before there is strong covariance between different bins
in $w_2(\theta)$. This is mostly due to the fact that the lowest
$l$ multipoles, and therefore larger scales, have more power than 
higher $l$ multipoles (e.g. see Fig. \ref{cl.eps}). As $w_2(\theta)$ is
dominated by large scale modes, the correlations at different scales
are strongly correlated. This is obviously worse in the \lcdm model than 
in the low-Q \lcdm, which has a lower quadrupole and octopole. 

It is therefore
essential that we estimate the covariance between different
bins in  $w_2(\theta)$. This can be calculated
 from the simulations by using the
following definition of the covariance matrix:
\bea{eq:covar}
C_{ij} &\equiv& 
\langle\Delta w_2(\theta_i)~\Delta w_2(\theta_j)\rangle\nonumber\\
&=& {1\over{N}} \sum_{L=1}^{N} \Delta w_2^L(\theta_i) \Delta w_2^L(\theta_j),
\eea
where
\be{eq:covar2}
\Delta w_2^L(\theta_i) \equiv w_2^L(\theta_i) - \widehat{w_2}(\theta_i).
\ee
Here $w_2^L(\theta_i)$ is the 2-point 
function measured in the $L$-th realization ($L=1\dots N$)  and $\widehat{w_2}(\theta_i)$ 
is the mean value for the $N$ realizations.
The case $i=j$ gives the error variance: $\sigma_i^2 \equiv C_{ii}$.

It is also useful to define the normalised covariance matrix (or correlation matrix), as:
\be{eq:ncovar}
\widetilde{C_{ij}} \equiv {C_{ij}\over{\sqrt{C_{ii}{C_{jj}}}}}
\ee
\begin{figure*}
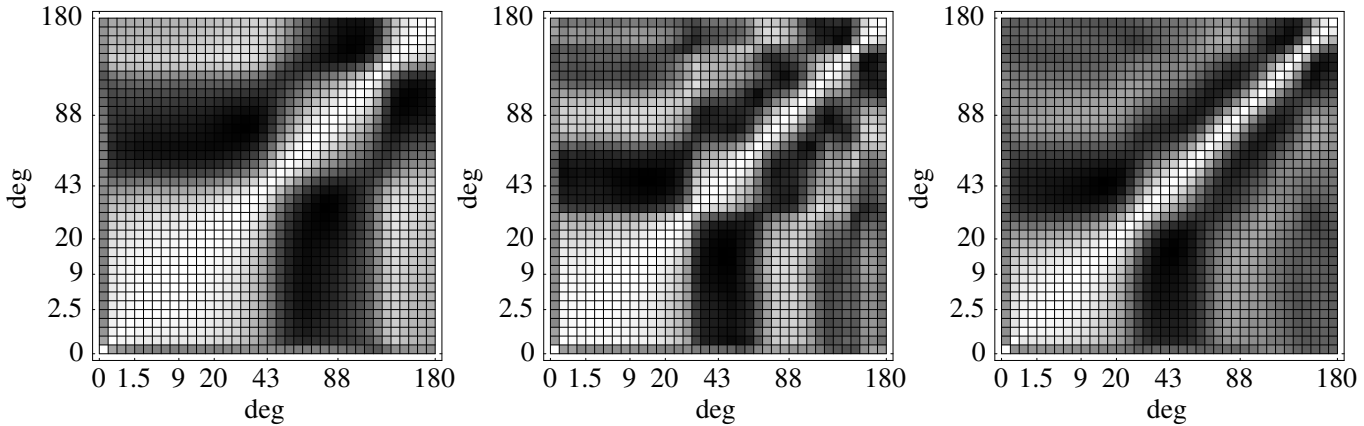

\vskip 2.5 truein
\includegraphics{nlcdm.ps}
\includegraphics{nlowq.ps}
\includegraphics{nlowq2.ps}
\caption
{Normalised covariance (or correlation) matrices between different 
angular bins in \w. Grey scale goes from -1 (black) to +1 (white). 
Left: \lcdm model. Middle: low-Q model. Right:
jackknife samples in low-Q model.}
\label{ncovar}
\end{figure*}

Figure \ref{ncovar} shows the normalised covariance in the \lcdm
(left) and low-Q model (middle). The covariance shown for the \lcdm
model corresponds to 40 realizations, but very similar pictures are
obtained for a 1000 realizations.  The covariance matrix is quite
different in these two cases as the \lcdm model shows correlations on
larger scales because of the higher quadrupole and octopole. Note in
particular the strong anticorrelation between angular bins in the
range 50-100 degrees with the bins on larger and smaller scales. This
is clearly an indication of very large variations in the low order
multipoles within different realizations (eg see Fig.\ref{w2low}).
Note also how the errors in the first bin (zero-lag in $w_2$) is
dominated by instrumental noise so that this bin is not correlated
with the rest of the bins (grey color corresponds to zero).

\subsection{Jackknife covariance}

We can also estimate the errors in $w_2(\theta)$
 from a single simulation, or from the real
sky with a variation of the jackknife error scheme proposed by
Scranton \etal (2001). This has the potential advantage of producing
an error estimate which is model independent. In the estimation,
the sample is first divided into $M$ separate regions on the sky, 
each of equal area. The analysis
is then performed $M$ times, on each occasion removing
 a different region. These are called
the (jackknife) subsamples, which we label $k=1 \dots M$. The estimated
statistical covariance for $w_{2}$ at scales $\theta_i$ and $\theta_j$
is then given by:
\bea{covarjack}
C_{ij}
&=& {M-1\over{M}} \sum_{k=1}^{M} \Delta w_{2}^k(\theta_i) \Delta
w_{2}^k(\theta_j) \\
& &\Delta w_{2}^k(\theta_i) \equiv w_{2}^k(\theta_i) - \widetilde{w_{2}}(\theta_i)
\eea
where $w_{2}^k(\theta_i)$ is the measure in the 
$k$-th subsample ($k=1
\dots M$)  and $\widetilde{w_{2}}(\theta_i)$ is the mean value for the $M$ subsamples.
The case $i=j$ gives the error variance. Note how, if we increase the number of
regions $M$, the jackknife subsamples are larger and each term in the sum is smaller. 
We typically take $M=16$ corresponding to dividing the sphere into 8 octants and each
octant in two (by longitude). We have checked that the 
resulting covariance gives a stable answer
 for different choices of shapes and $M$
from $M=8$ to $M=32$. 

The mean normalised jackknife covariance shown in the right panel of
Fig. \ref{ncovar} is in very good agreement with the sample covariance 
shown in the middle panel. The jackknife covariance appears smoother
because it is the mean of 20x16 jackknife samples, while the sample 
covariance is the mean
of only 20 samples. Other than this, the jackknife matrix correctly captures all the 
relevant correlation information present in the ensemble.
The corresponding diagonal errors are shown in Fig. \ref{ew2vmap}.

In general, one would not expect such a good agreement, because
the jackknife method can not account for variations on scales larger than
the jackknife samples. In our case, however, by construction there is little power on
the largest scales (quadrupole and octopole) in the low-Q model. This explains
the good performance of the jackknife method.


\subsection{$\chi^2$ test}

In order to quantify how well a model fits the data
we use a $\chi^2$ test.
 Since the bins of the \w are correlated, however, the simple 
$\chi^2$-test is not valid  but must be modified by using the covariance matrix.
The value of the $\chi^2$ is given by
\be{eq:chi}
\chi^2 = \sum_{i,j=1}^{N} \Delta_i ~ C_{ij}^{-1} ~ \Delta_j,
\ee
where $\Delta_i \equiv w_2^O(\theta_i) - w_2^M(\theta_i)$
is the difference between the "observation" $O$ and the
model $M$.
In terms of the reduced covariance matrix, this is
\be{eq:chired}
\chi^2 = \sum_{i,j=1}^{N} \Big({\Delta_i\over\sigma_i}\Big) ~ \widetilde{C}_{ij}^{-1} ~ 
\Big({\Delta_j\over\sigma_j}\Big).
\ee

\subsection{Singular Value Decomposition}

In using real world data, one must worry about degeneracies in
the covariance matrix due to an over-determined system and noise. 
If the rows (or columns) of $C_{ij}$ are not independent, the
determinant of the matrix is zero and there is no inverse. With real data,
the determinant of $C_{ij}$ will not typically be zero but the matrix
can have very small eigenvalues which then lead to artificially
large eigenvalues of the inverse matrix. 

In order to eliminate the degeneracies in the covariance matrix, we 
perform a Singular Value Decomposition (SVD) of the matrix,
\be{eq:svd}
\widetilde{C_{ij}}=(U_{ik})^\dagger W_{kl}V_{lj},
\ee
where $W_{kl}$ is a diagonal matrix with the singular values on the diagonal.
By doing the decomposition, we can choose the number of modes 
we wish to include in our $\chi^2$ by effectively setting the corresponding 
inverses of the small singular values to zero.

In practice we find that the SVD is only required when we use a small
number of simulations to estimate the covariance matrix. With more than 
100 simulations we find that the covariance matrix inversion is
stable and one then gets more accurate $\chi^2$ values.


\subsection{Recovering $C_l$ from $w_2(\theta)$}

\begin{figure}
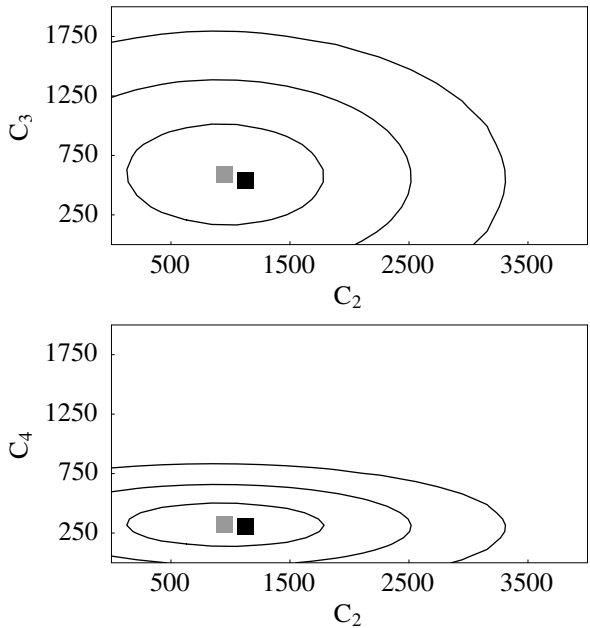

\vskip 3.4 truein
\includegraphics{lcdm23.ps}
\includegraphics{lcdm24.ps}
\caption{Recovered values of $(C_2,C_3)$ (upper panel) and $(C_2,C_4)$ 
(lower panel) from the \lcdm model. The black box indicates the
values used to generate \w and the gray box the recovered value. 
Confidence contours are plotted at 10, 68 and 98\% of the minimum $\chi^2$ value.}
\label{lcdmcls}
\end{figure}

\begin{figure}
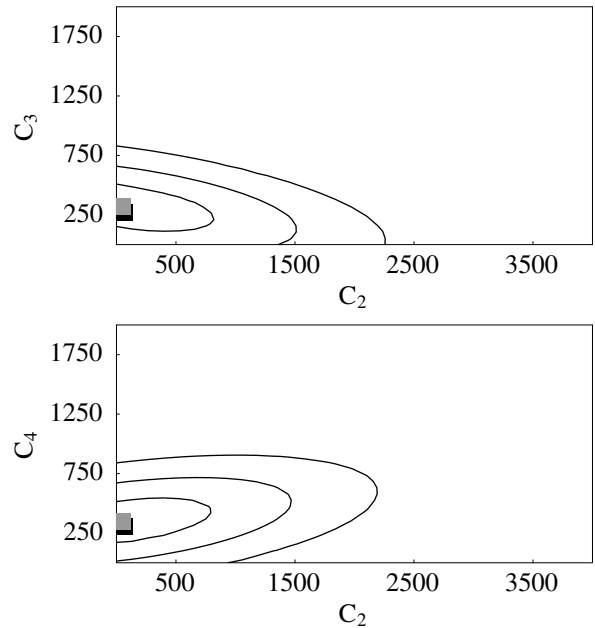

\vskip 3.4 truein
\includegraphics{lowq23.ps}
\includegraphics{lowq24.ps}
\caption{Recovered values of $(C_2,C_3)$ (upper panel) and $(C_2,C_4)$ 
(lower panel) from the low-Q model. The black box indicates the
values used to generate \w and the gray box the recovered value.
Confidence contours are plotted at 10, 68 and 98\% of the minimum $\chi^2$ value.}
\label{lowqcls}
\end{figure}

Using the SVD of the covariance matrix we can 
recover $C_l$s from $w_2(\theta)$ by requiring a minimum $\chi^2$
(see Szapudi \etal 2001 for a different approach). 
To test the method, we have recovered the low $C_l$s from the \lcdm and
low-Q modes. These are shown in Figures  Fig. \ref{lcdmcls} and  Fig. \ref{lowqcls} respectively.
The confidence contours are plotted relative to the minimum $\chi^2$
value.
As it is clear from the figures, the method accurately reproduces the correct
values of the $C_l$s. We have also checked that this method is robust 
and insensitive to the number of modes used in the calculation.


\section{Results from WMAP}

\subsection{Estimation of $w_2(\theta)$ in WMAP}

\begin{figure}
\vskip 3 truein
\includegraphics{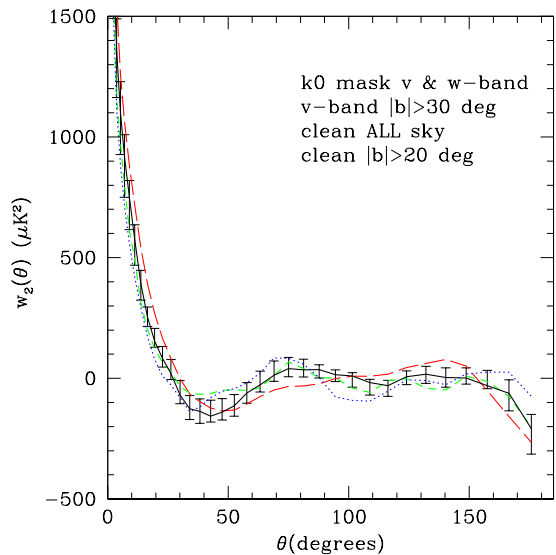}
\caption{Error bars represent the 2-point angular correlation and 1-sigma
jackknife dispersion $w_2$ in the WMAP V-band.
 The continuous line, passing through the error-bars, shows  $w_2$ in the WMAP W-band.
These two cases have the kp0 mask, while the dotted line excludes all
the Galactic plane with $b>30$ deg. The long  and short  dashed lines
correspond to the foreground  clean map of Tegmark etal (2003) with and without
a Galactic cut $|b|>20$.}
\label{w2wvclean}
\end{figure}

Figure \ref{w2wvclean} shows different estimations of $w_2(\theta)$ 
from the WMAP data. 
These different estimates gives an idea of the systematics involved,
which are related mostly to possible Galactic and foreground contamination
in the maps (see Bennet \etal 2003, Tegmark \etal 2003 for more details).
Also note how the jackknife errors (from the dispersion in the V-band
estimates from different regions in the sky) roughly include the variations in these
different estimations. Thus we will use V-band estimates with jackknife errors
as representative of WMAP sampling variance and systematic uncertanties.

Note how our estimates of \w agree well with that presented by the WMAP
collaboration (e.g. compare to Fig.13 in Bennett et al. 2003).

\subsection{Comparison of $w_2(\theta)$ to simulations}

\begin{figure}
\vskip 3 truein
\includegraphics{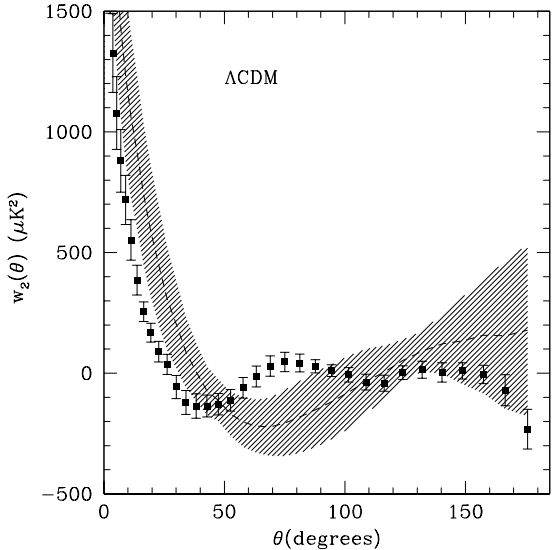}
\caption
{The 2-point angular correlation $w_2$ in the WMAP V-band
data (points with jackknife error bars) compared to the 
 $\Lambda$CDM simulations (shaded region corresponds to
 1-sigma error dispersion). In both cases we use the kp0 
Galactic mask.}
\label{w2vmapwm}
\end{figure}

\begin{figure}
\vskip 3 truein
\includegraphics{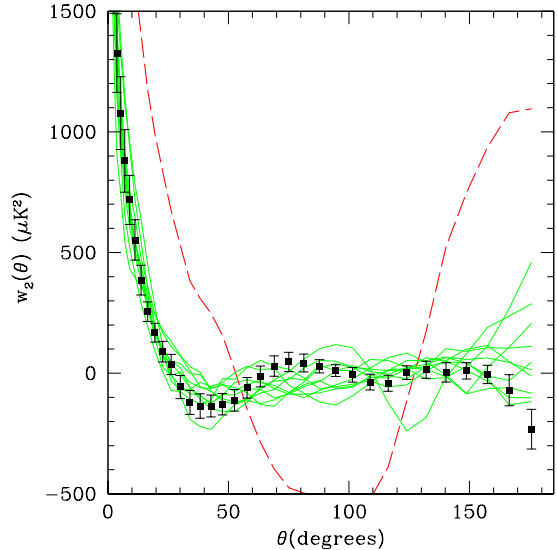}
\caption
{The 2-point angular correlation $w_2$ in the WMAP V-band
data (points with jackknife error bars) compared to 10 of the
 $\Lambda$CDM simulations (continuous lines) which happens
 to have a low quadrupole
 and one (dashed line) with a high quadrupole. These are all taken out
 of the 100 realisations, whose dispersion is 
 shown as the shaded region in Fig.\ref{w2vmapwm}.}
\label{w2vmapwm2}
\end{figure}

Figure \ref{w2vmapwm} compares the WMAP $w_2(\theta)$ 
estimation with the 1-sigma dispersion in the corresponding
\lcdm simulations. At first sight one might conclude that a \lcdm model is ruled out
to a high significance. 
It should be remembered, however, 
 that there is a strong covariance in the $w_2(\theta)$
estimation (see Fig.\ref{ncovar}). This point
is illustrated in Figure \ref{w2vmapwm2} which
shows 10 of the \lcdm realizations that happens
 to have a low quadrupole and are compared with one 
 that happen to have a high
 quadrupole (all chosen from the first 100 \lcdm simulations).
Note how systematics effects in WMAP (which are well represented by
the jackknife errors) are comparable to the differences between these
10 \lcdm simulations and the mean WMAP estimation.

\begin{figure}
\vskip 3.7 truein
\includegraphics{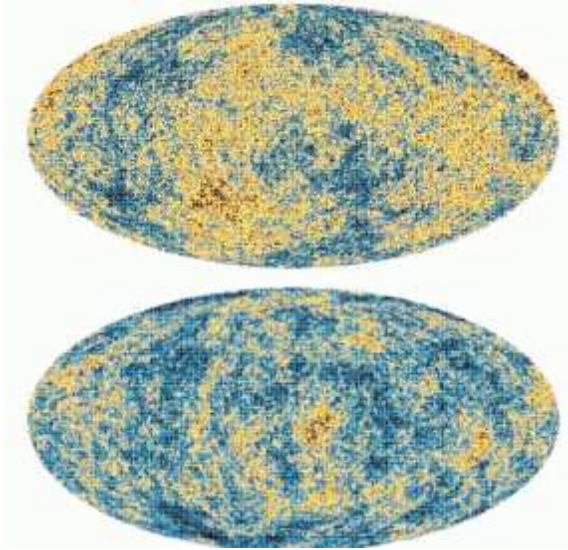}
\caption
{Two of the \lcdm simulations, with average (top) and low (bottom) quadrupole.}
\label{wmaps}
\end{figure}

These very obvious differences in the lowest multipoles in some
of the realizations are also apparent 
in the all-sky simulated maps.
Figure \ref{wmaps} illustrates this point by showing
one of the low quadrupole \lcdm realizations  and   a \lcdm realization 
with the average value of the quadrupole.  

\begin{figure}
\vskip 3 truein
\includegraphics{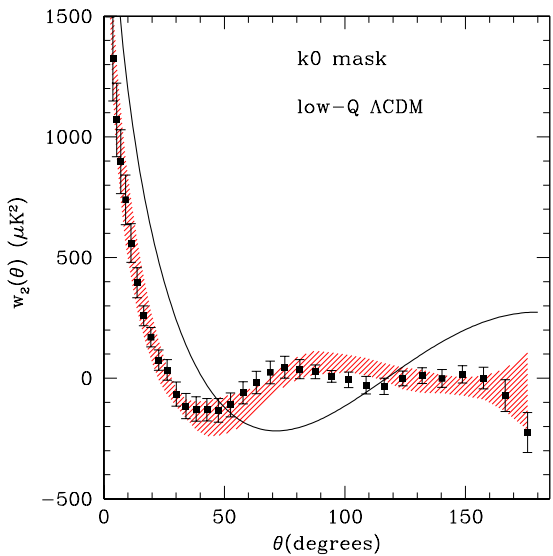}
\caption
{The 2-point angular correlation $w_2$ in the WMAP V-band
data (points with jackknife error bars) compared to the low-Q
simulations (\lcdm model with  zero quadrupole, $c_2=0$, and 1/2
the octopole $c_3/2$). The continuous line shows the \lcdm model
predictions.}
\label{w2vmanoq}
\end{figure}

In Figure \ref{w2vmanoq} we show the comparison of $w_2(\theta)$
with the low-Q model which is apparently in 
much better agreement with WMAP.

\subsection{Covariance Matrix}

\begin{figure}
\vskip 2.5 truein
\includegraphics{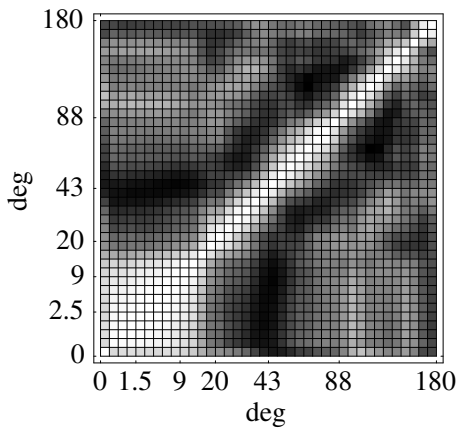}
\caption
{Normalised covariance matrix from WMAP jackknife samples, to be compared to
Fig. \ref{ncovar}.}
\label{ncovar1}
\end{figure}

To quantify these differences more accurately
we need to calculate the covariance matrices of the models.
We can also estimate the covariance matrix of the data using the 
jackknife method, Eq. [\ref{covarjack}].
Figure \ref{ncovar1} shows the normalised covariance matrix estimation,
which is more similar to the low-Q model than to the \lcdm model
shown in Fig. \ref{ncovar}. Fig. \ref{ew2vmap} compares the
diagonal errors with N=8 and N=32 jackknifes subsamples (continuous
lines) with the errors in the low-q and \lcdm models.

\subsection{$\chi^2$ tests of the WMAP data}

\begin{table}
\caption{Likelihood of WMAP data in different models.
First column shows the model we are comparing to, e.g. 
\ \lcdm denotes that we are comparing to the mean \w \ of the \lcdm
simulations. The second column indicates the model from which we take
the covariance matrix. 
The fourth column indicates how many singular values we use in 
calculating the $\chi^2$ and the fifth column is the probability
of the fit.
\label{tab:chisq}}
\begin{center}
\begin{tabular}{ccccc} \hline \hline
\smallskip
Model & Errors & $\chi^2$ & Modes & P\\
\lcdm & WMAP & 200 & 7 & 0\\
\lcdm & \lcdm & 3.5 & 5 & 0.62\\
low-Q & WMAP & 9.6 & 7 & 0.21\\
low-Q & low-Q (true)& 4.7 & 5 & 0.45\\
low-Q & low-Q (jackknife) & 4.1 & 6 & 0.66\\
\hline
\end{tabular}
\end{center}
\noindent
\end{table}

\begin{table}
\caption{Likelihood of WMAP data with a galaxy cut in different models
\label{tab:chisq30}}
\begin{center}
\begin{tabular}{ccccc} \hline \hline
\smallskip
Model & Errors & $\chi^2$ & Modes & P\\
\lcdm 30 & WMAP 30 & 191 & 7 & 0\\
\lcdm 30 & \lcdm 30 & 5.7 & 6 & 0.46\\
low-Q 30 & WMAP 30 & 14 & 7 & 0.06\\
low-Q 30 & low-Q 30 & 5.2 & 7 & 0.64\\
\hline
\end{tabular}
\end{center}
\noindent
\end{table}

Using the method described in the previous section, 
we  compare the WMAP \w to different models in 
Table \ref{tab:chisq}. The second row shows how
considering the WMAP data as a realization of the \lcdm model
gives a very good fit (62\% C.L.).
However, considering the WMAP jackknife errors\footnote{A similar result
is obtained using low-Q \lcdm true errors.}  (i.e. first row in Table\,2) 
leads to a vanishing probability 
for the \lcdm model, as was already hinted at in  Fig.\ref{w2vmapwm}.

On the other hand, the WMAP data provides a resonable fit
to the low-Q
models, even if we use the WMAP errors, as can be seen from the 
third, fourth and fifth row of Table \ref{tab:chisq30}.
The WMAP data clearly
has a low quadrupole compared to \lcdm as was already suggested by Figure 13. 

Considering only the WMAP data at $|b|>30$ deg,
 and comparing the resulting $w_2(\theta)$
to the models, give similar results as can be seen in Table 
\ref{tab:chisq30}. 
This is in qualitative agreement  with that 
found by COBE (Kogut \etal 1996).

\begin{figure}
\vskip 2.5 truein
\includegraphics{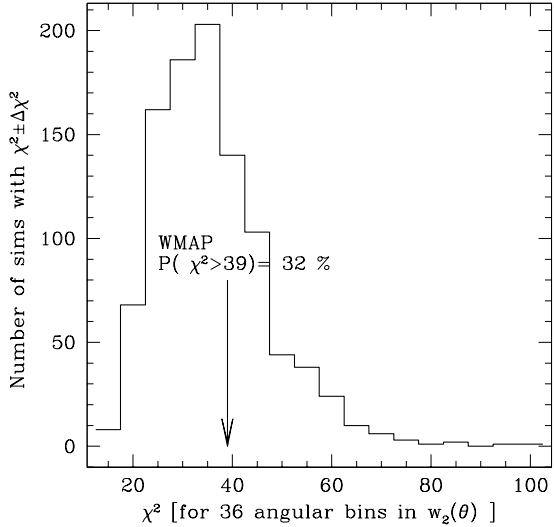}
\caption
{Number of \lcdm simulations
with given values of $\chi^2$ in Eq.\ref{eq:chi} using the 
covariance matrix from the same \lcdm simulations. 
The same test for WMAP gives: $\chi^2 \simeq 39$ (shown
by the arrow), which is below $32\%$ of the simulations.}
\label{chi2hist}
\end{figure}

The above results are obtained with a relatively small number of
simulations (40 for \lcdm and 20 for low-Q model) and also a small
number of jackknife subsamples (16). Thus, although these estimates probably
give the right order of magnitude for the calculation we can not
expect them to be very accurate.
 We next quantify what are the errors involved by repeating the \lcdm analysis 
with an increasing  number of simulations, from $N=100$ to $N=2000$,
to calculate the covariance matrix. 
We can also check with these additional
simulations  if the $\chi^2$ distribution gives
an accurate representation of the probabilities. This is illustrated
in Fig.\ref{chi2hist}, which shows the histogram of $\chi^2$ values 
obtained for each \lcdm realization using a
 direct inversion of the covariance matrix (no SVD), estimated
from $N=1000$ \lcdm realizations. The histogram
matches well the probabilities given by the $\chi^2$ distribution. 
Moreover we find that, other than Poisson errors, the values are
quite insensitive to the total number $N$ of simulations used. 
The probability of finding a \lcdm simulation with a $\chi^2$
value larger than WMAP seems to converge to $32\%$. This number
compares well to the crude $62\%$ estimate we found in Table \ref{tab:chisq}
using SVD over  40 simulations. If we used 20 or 100 simulations with SVD we 
find intermediate results. We have also estimated the $\chi^2$ values
for  $N=2000$ \lcdm simulations  using the WMAP jacknife covariance
matrix. We find that none of the  $N=2000$ \lcdm simulations have
a $\chi^2$ smaller than WMAP (the smaller value in the 2000 simulations
is $\chi^2=62$ for 36 bins in \w). Thus the probability for
\lcdm given the WMAP errors is $P< 0.05$\%, in good agreement
with first entry in Table \ref{tab:chisq}. 

\subsection{Recovered $C_l$ from $w_2(\theta)$}

Using the previously discussed method (see \S 2.11)
 of calculating the $\chi^2$ with
the SVD, we can recover
the $C_l$s from the data by requiring that $\chi^2$ has a minimum.
Fitting three multipoles ($C_2,\ C_3$ and $C_4$) of a \lcdm model 
to the WMAP data and using the \lcdm errors yields
\be{be:2minpoint}
C_2=-7\ C_3=190\ C_4=226.
\ee
With the jacknife WMAP errors we get
\be{be:2minpoint2}
C_2=123\ C_3=217\ C_4=212.
\ee
 Comparing with Eq. \ref{eq:cls}, we see that the best fit value
has a negligible quadrupole and an octopole that is roughly half the value
of the \lcdm model. Even the $C_4$ is somewhat lower than the \lcdm value.
The low-Q model is hence a better approximation to the \w measured from 
the WMAP data.

\begin{figure}
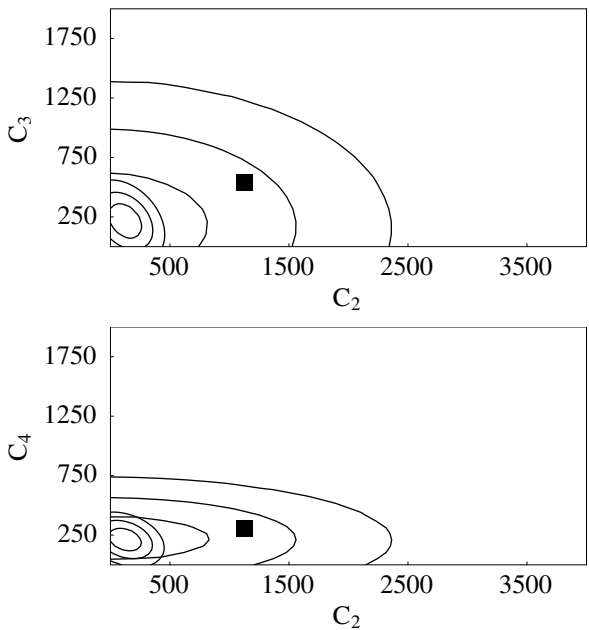

\vskip 3.4 truein
\includegraphics{wmap23.ps}
\includegraphics{wmap24.ps}
\caption{Confidence plot of $(C_2, C_3)$ (upper panel) and $(C_2, C_4)$ 
(lower panel) recovered from the \lcdm model with WMAP (inner contours) and
\lcdm (outer contours) errors.
The contours, in both cases,  are at the $0.1,\ 0.68$ and $0.98$ confidence levels. 
The filled square shows the location of the theoretical \lcdm values in both cases.}
\label{wmapconts}
\end{figure}

The confidence contours of the $C_l$s are plotted
in Fig. \ref{wmapconts} for the WMAP errors for
the two sets of pairs $(C_2,C_3)$ and $(C_2,C_4)$.
The figure fully confirms our expectations: the WMAP data is in good
agreement with \lcdm
when WMAP data is considered a realization of the \lcdm model (i.e. using
\lcdm errors). The WMAP data is not in agreement with \lcdm however
 when using WMAP errors.

From the confidence plots we can recover the $68\%$ C. L. 
(or $\sim 1$-sigma) limits for the $C_l$s by 
considering the projections of the contours
on the axes. When no model  
is assumed, i.e. we use WMAP errors,
we find (in $\mu K^2$):
\bea{eq:wmapcls}
C_2 & = & 123 \pm 233\nonumber\\
C_3 & = & 217 \pm 241\nonumber\\
C_4 & = & 212 \pm 162,
\eea
which are shown as closed triangles in Fig.\ref{cl.eps}.
With a \lcdm covariance matrix the errors are much larger (in $\mu K^2$),
\bea{eq:wmapcls2}
C_2 & = & -7 \pm 1568\nonumber\\
C_3 & = & 190 \pm 799\nonumber\\
C_4 & = & 226 \pm 339,
\eea
Compared to Eq.\ref{eq:wmap2003cls}, our errors are about 
a factor of two larger.

\begin{figure*}
\vskip 2.2 truein
\includegraphics{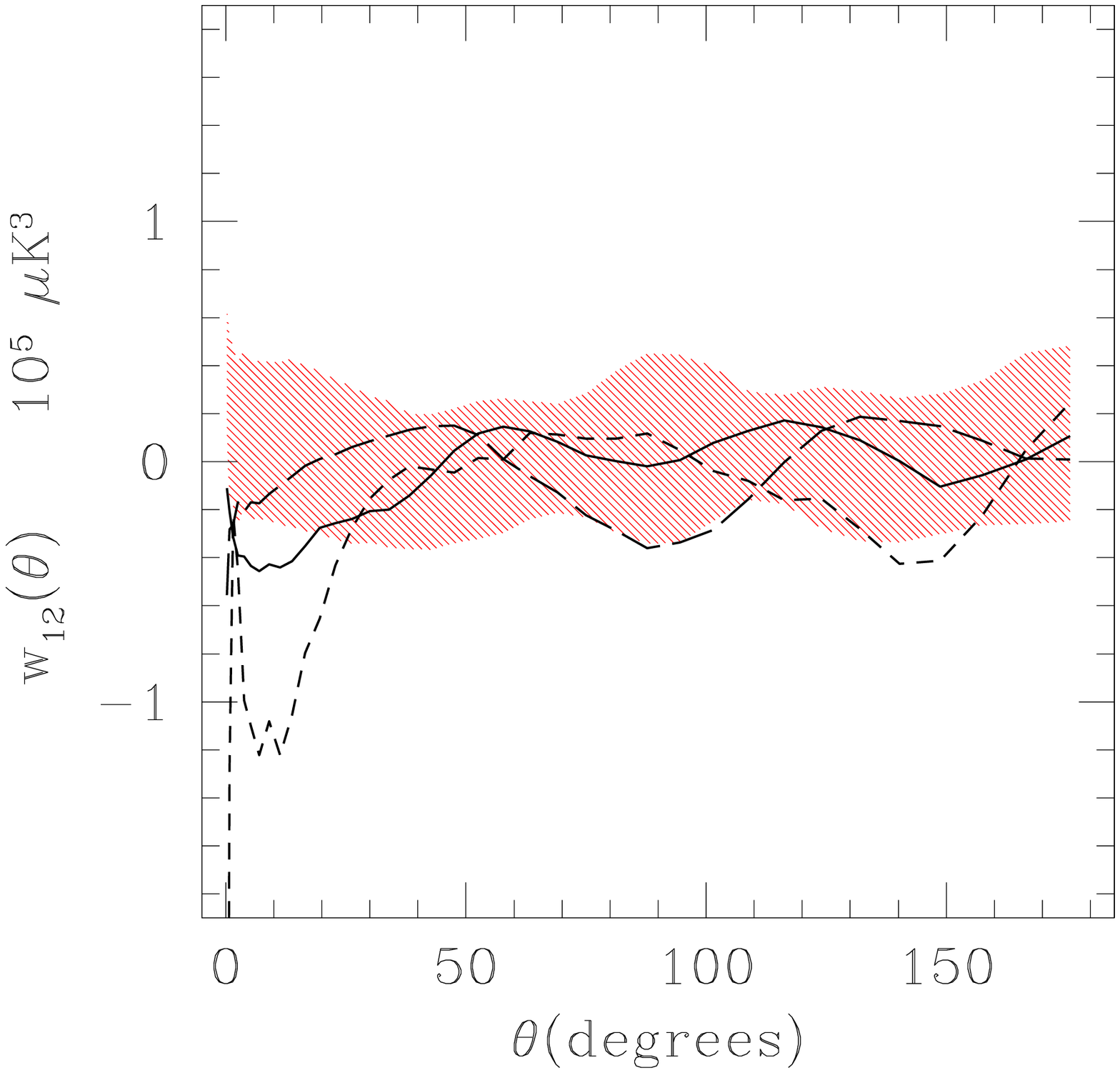}
\includegraphics{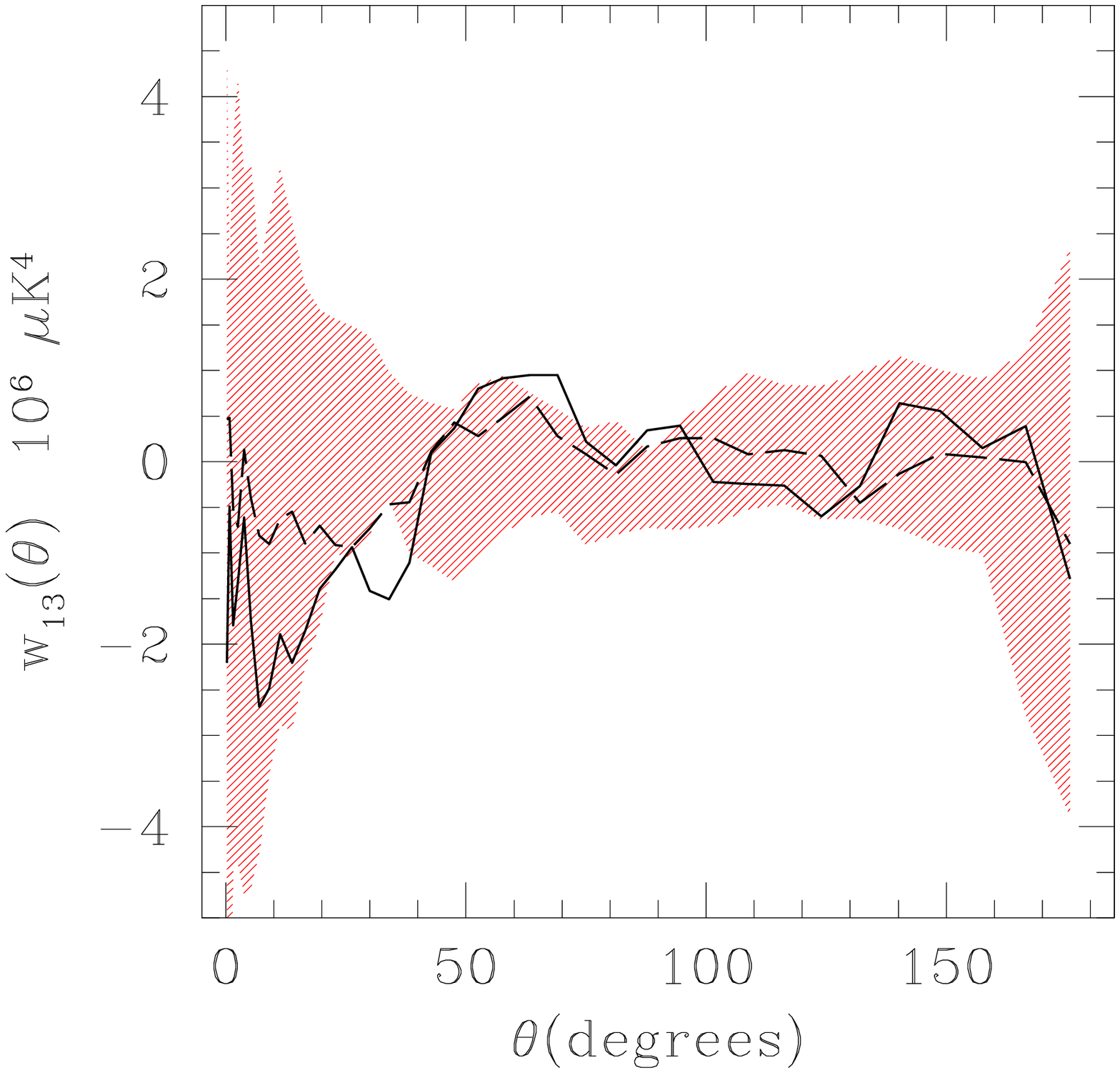}
\includegraphics{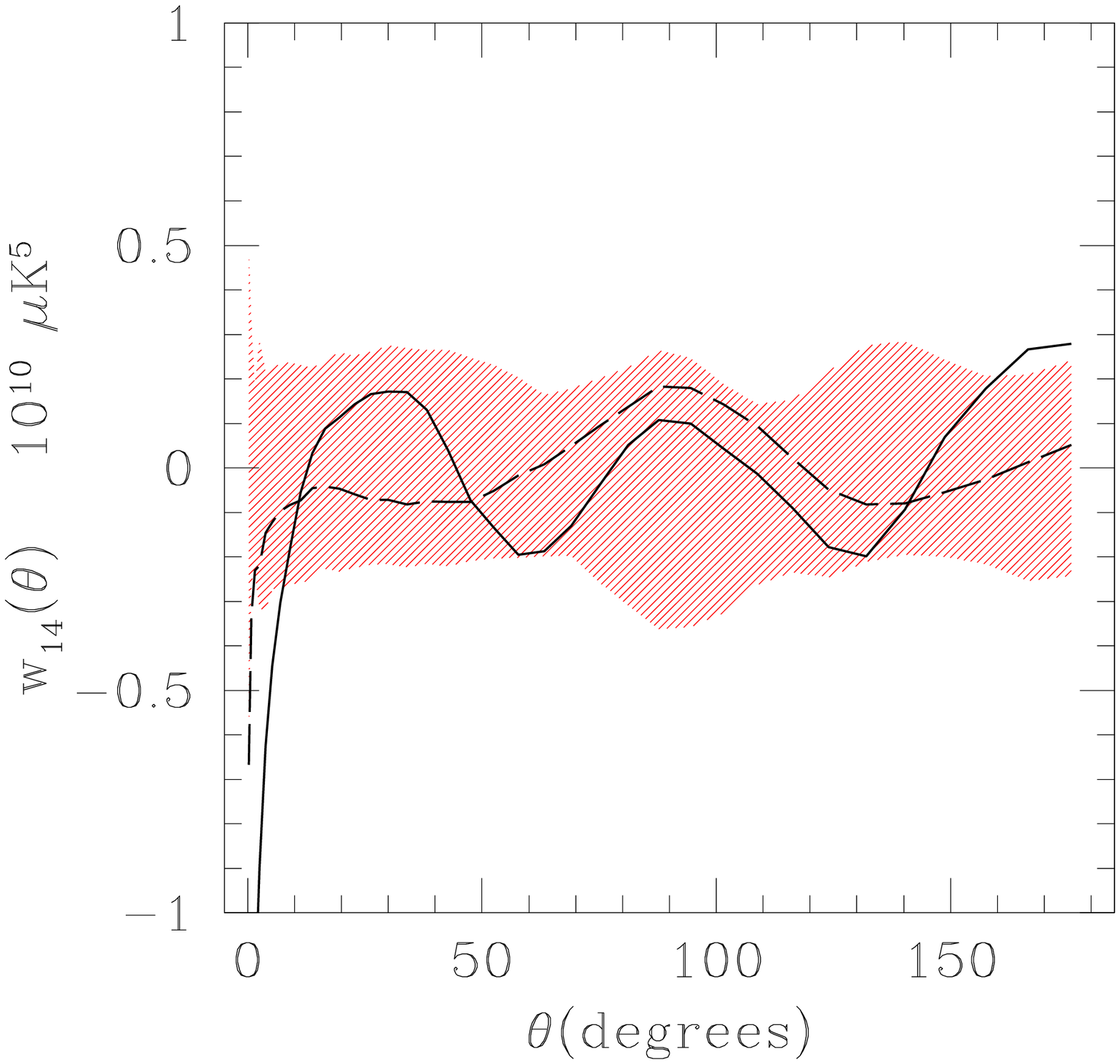}
\caption
{Higher-order 2-pt moments in WMAP V-band (continuous line) and WMAP foreground 
clean map (long-dashed line) as compared with the dispersion in the corresponding 
measurements in the low-Q \lcdm simulations (shaded regions). 
The short-dashed line in the left-hand panel shows $w_{12}$ 
from the full combined clean map, ie 
without the Galactic cut at $|b|<20$.}
\label{w14}
\end{figure*}

\subsection{Higher-order moments}

We also present  the higher-order moments of the 2-point
correlations in WMAP. This provides a consistency test for the Gaussian
hypothesis, which is implied in the comparison with simulations.

The higher-order moments of the 2-point function can be defined as:
\be{kpq}
k_{pq}(\theta) \equiv \vev{ \Delta T^p({\bf\theta_1}) \Delta T^q({\bf\theta_2}) }
\ee
where $p=q=1$ corresponds to the 2-point function in Eq.[\ref{w2def}],
i.e. $k_{11}(\theta)=w_2(\theta)$ and
$p=0$ gives the 1-point moments, e.g. $k_{02}=k_{11}(0)$ is the variance
and $k_{03}=k_{12}(0)$ is the skewness.
It is then useful to define 2-point cumulants as {\it connected} moments, 
subtracting the lower-order contributions
(e.g. see Gazta\~naga, Fosalba \& Croft 2002)
:
\bea{wpq}
w_{11}(\theta) &\equiv & w_2(\theta) \equiv  k_{11}(\theta) \\
w_{12}(\theta) &\equiv & k_{12}(\theta) \\
w_{13}(\theta) &\equiv & k_{13}(\theta) -3  k_{11}(0)~k_{11}(\theta)\\
w_{14}(\theta) &\equiv & k_{14}(\theta) -6 k_{11}(0)~k_{12}(\theta)
-4 k_{12}(0)~k_{11}(\theta)
\eea
For a Gaussian field we expect $w_{pq}(\theta)=0$.

It is straightforward to estimate these higher-order moments in the same
way in which we have calculated the 2-point function:
\be{wpqest}
k_{pq}(\theta) = {\sum_{i,j} \Delta_i^p \Delta_j^q ~w_i^p~w_j^q 
\over{\sum_{i,j} w_i^p~w_j^q}}.
\ee
Here, as in calculating the 2-point function, uniform weights, $w_i=1$,
are used. We have estimated the higher order moments 
as well as the covariance matrix both from WMAP and simulated data. 

Figure \ref{w14} shows $w_{12}$, $w_{13}$ and $w_{14}$ for the WMAP
V-band data and the combined foreground clean map in Tegmark et
al. (2003) with a Galactic cut excluding $|b|<20$ (long dashed line).
Also shown are the 68\% CL in the dispersion from the low-Q \lcdm
simulations. Both WMAP and the combined map with $|b|>20$ agree well
with the Gaussian hypothesis $w_{pq}=0$ at high significance.
 
For the 3rd order moment, $w_{12}$, we also show the combined
clean map without the Galactic cut (short-dashed line).
This result is incompatible 
with the Gaussian low-Q \lcdm simulations (and covariance matrix) on
small  angular scales, $\theta <20$ deg. This is in qualitative
agreement to what Chiang \etal (2003) 
have recently found using phase correlators over the very same map.
Nevertheless, we note how this apparent
 discrepancy goes away when we exclude the $|b|<20$  
Galactic region.
Thus, this non-Gaussian feature seems to be related to residual
Galaxy contamination, which is visually apparent in the image of the
combined clean map (e.g. see Fig. 1 in Tegmark et al. 2003).

Our best constraints on non-Gaussian features come from
the 4th-order statistic,
$w_{13}$, because in this case the pixel noise at two different points
 would tend to cancel out, as in the 2nd-order moment $w_2$.  This is not the 
case for 
the 3rd and 5th-order moments, $w_{12}$ or $w_{13}$,  where pixel noise
is added in quadrature. One needs to  consider  multi-point correlations 
to reduce the noise contribution and get better constraints 
on non-Gaussian models, which is beyond the scope of this paper
(see Hinshaw \etal 1995, Komatsu \etal 2003). The point here
is that the higher-order 2-point anisotropies in WMAP are perfectly compatible
with the Gaussian hypothesis, given the pixel noise. We find
similar results up to 8th-order moments. 


\section{Conclusions}

In this paper we have studied the lowest-order multipoles in 
the CMB from recent WMAP data. This work is motivated by the
fact that the low order multipoles, and especially the quadrupole as
measured in harmonic space,  show less
power than that  expected in the standard \lcdm model
(e.g. see Bennett \etal 2003; Spergel \etal 2003; 
Peiris \etals 2003). Instead of looking at the power spectrum, we study the WMAP data
by using the two-point angular correlation function, \w, as a tool
to estimate the multipoles. The main
advantage is that \w is very sensitive to the low order multipoles
The main disadvantage is that different bins of \w are highly correlated
so a covariance analysis is required.

We have simulated a set of V-band maps with different random phases
using the best-fit WMAP power spectra and pixel errors, and calculate
the 2-point function for each model. Using this set of simulations we
calculate the mean value and variance at each bin as well as the
covariance matrix. The errors and the covariance matrix of the WMAP
data is calculated by using the jacknife method. A \lcdm model with a
zero quadrupole and a low octopole is also simulated and used to test
the reliability of the jacknife errors.

Using a $\chi^2$ test with a
covariance matrix we calculate the significance of the WMAP data in
different models. It is found that indeed, the WMAP data with the WMAP
errors does not fit the \lcdm model and is strongly disfavoured.
Instead, the WMAP data fits well the low-Q model with a vanishing
quadrupole.  However, taking a different point of view and considering
the WMAP data as a sample realization of the simulated \lcdm maps, we
find that the WMAP results are fully consistent with the \lcdm
model. From this viewpoint, our sky just happens to have a low
quadrupole and is just as likely to have a large quadrupole as
demonstrated by the different realizations in Figure \ref{wmaps}.

Our results are in apparent contradiction with the harmonic space
analysis presented by Spergel \etal (2003), which gives WMAP a
probability as low as $0.15 \%$ of being a realization of the \lcdm
model.  In contrast we estimate this probability to be $32\%$
comparing \w in WMAP to \w in 1000 \lcdm realizations (see
Fig.\ref{chi2hist}).  Our probability is closer, but still larger than
the $5\%$ value obtained from harmonic analysis by Tegmark etal
(2003)\footnote{While this paper was being referee, we became aware of
another calculation by Efstathiou (2003b) who find intermediate
probabilities of $1-2\%$, also using direct harmonic analysis.}  Why
are all these estimates so different? To answer this question we
conduct a further test by repeating our calculation using a different
galactic mask.  We use a sharp cut in Galactic latitude $|b|$ of about
$\pm 20$ degrees, excluding the same number of pixels as the kp0
mask. We then recalculate the covariance matrix and its inversion with
this new Galactic cut over the same 1000 \lcdm realizations.  
We repeat the $\chi^2$ analysis (in Fig.\ref{chi2hist})
for WMAP and the 1000 simulations with this new covariance matrix.
We find that only $1\%$ of the simulations with this Galactic cut have a
$\chi^2$ larger than the one measured for WMAP.  Thus the final
probability depends strongly on the shape of the mask, and therefore
on how the mask is accounted for in the comparison between the models
and observations.  This is not surprising because the large scale
modes that determine the shape of \w (eg Fig.\ref{w2low})
extend over angular regions that are comparable to the mask. To
recover the low order multipoles from masked data using harmonic
decomposition, one needs to invert
the convolution with the harmonic image of the mask. This is a
non-linear transformation that is likely to introduce biases in the
recovered low multipoles.  We speculate that the origin of the above
discrepancies with the harmonic analysis could depend on how the mask
is taken into account on the different estimators. Because the \w
estimator is independant of the mask we believe that our $\simeq 30\%$
probability value is reliable.

One can also recover the values of the low multipoles from the \w
WMAP estimations.  The validity of the method is demonstrated by
recovering known $C_l$s from a set of simulated data.  Again, we find
that the best fit model has a low quadrupole as well as a small
octopole.  Confidence levels in the $C_l$ plane point to a similar
conclusion: the WMAP data fits the \lcdm model, but only with \lcdm
priors.  We can also reach a similar conclusion by asking how many of
the first 100 \lcdm simulations show a 2-point function that "looks"
closer to the WMAP data than to \lcdm $w_2$ (e.g. see
Figure\ref{w2vmapwm2}).  There is a clear distinction of such cases,
which is even apparent by a direct inspection of the pixel-map images
(see Figure \ref{wmaps}).  We find at least 15 similar examples of low
quadrupole amongst the first 100 simulations, confirming our more
quantitative analysis above.

We have also considered the presence of possible non-Gaussianities in
the WMAP data by calculating higher-order moments of the 2-point
correlation. We find that both the WMAP and combined foreground clean
map of Tegmark et al. (2003) at $|b|>20$ are in excellent agreement
with the Gaussian hypothesis. It is also shown how the full-sky
foreground clean map (i.e. including $|b|<20$ ) shows a non-Gaussian
feature at angular scales $\theta < 20$ degrees, indicating a possible
residual galaxy contamination in the map.

In summary, the low value of the WMAP quadrupole is confirmed here by
looking at the two-point angular correlation function, \w, which is
shown to be a well-suited method for the study of low multipoles in
the CMB data.  This does not necessarily indicate a need for new
physics or non-standard cosmology, since we also show the WMAP \w fits
well a standard \lcdm cosmology if one considers our universe as a
realization of the set of possible \lcdm outcomes. We stress
nevertheless, that by considering the errors from the WMAP data alone,
i.e. with the jacknife errors, \w in WMAP is strongly inconsistent
with the mean \lcdm model and, under such an assumption, new ideas to
explain the discrepancy could well be in order.  Finally, we have
demonstrated that our conclusions are robust when we exclude data with
Galactic latitudes $|b|<30$, which indicates that the information (or
lack of information) in the Galactic plane is not responsible for the
missing large scale power.

\noindent
\section*{Acknowledgements} 
EG acknowledged support from INAOE, the Spanish Ministerio de Ciencia y
Tecnologia, project AYA2002-00850, EC-FEDER funding and
supercomputing center at CEPBA and CESCA/C4, where
part of these calculations were done. 
JW  acknowledged support from a scholarship at INAOE and
thanks Eric Hivon for providing assistance during the installation of
HEALPix. TM is grateful to the Academy of Finland for financial
support (grant no. 79447) during the completion of this work.
EG and TM are grateful to the Centre Especial de Recerca en 
Astrofisica, Fisica de Particules i Cosmologia
(C.E.R.) de la Universitat de Barcelona and IEEC/CSIC
for their support. DH and JW are grateful to financial support from 
CONACYT grant 39953-F.


\label{lastpage}

\begin{thebibliography}{}

\bibitem[]{BWMAP} Bennett C.L. \etal, 2003, preprint (astro-ph/0302207).

\bibitem[]{BONa} Bond J.R. \& Efstathiou G., 1987, MNRAS, 226, 655.


\bibitem[]{Chang} Chiang, L-Y., Naselsky, P.D., Verkhodanov, O.V., Way, M.J.
astro-ph/0303643

\bibitem[Contaldi et al. 2003]{conta}{Contaldi C.R., Peloso M., 
Kofman L., Linde A., 2003, preprint (astro-ph/0303636).}

\bibitem[]{Gazta} Gazta\~naga, E, Fosalba, P., Croft, R.A.C.
MNRAS, 331, 13.


\bibitem[Efstathiou 2003a]{efs03}
{Efstathiou G., 2003a, preprint (astro-ph/0303127).}

\bibitem[Efstathiou 2003b]{efs03b}
{Efstathiou G., 2003b, preprint (astro-ph/0306431).}


\bibitem[]{GHW99}
G\'orski, K. M., Hivon, E., \& Wandelt, B. D., 1999, in Proceedings of the
MPA/ESO Cosmology Conference "Evolution of Large-Scale Structure", eds. A.J. Banday, R.S. Sheth and L. Da Costa, PrintPartners
Ipskamp, NL, pp. 37-42 (also astro-ph/9812350)

\bibitem[]{Hin96a}
Hinshaw G., Banday A.J., Bennett C.L., Gorski K.M., Kogut A., 
1995, ApJ, 446, L67.

\bibitem[]{Hin96b}
Hinshaw G., Banday A.J., Bennett C.L., Gorski K.M., Kogut A.,  Smoot G.F.,
Wright E.L., 1996, ApJ, 464, L17.

\bibitem[]{Hin96c}
Hinshaw G., Banday A.J., Bennett C.L., Gorski K.M., Kogut A.,  
Lineweaver, C,H., Smoot G.F., Wright E.L., 1996, ApJ, 464, L25.

\bibitem[]{Kog96a}
 Kogut A.,  Banday A.J., Bennett C.L., Gorski K.M., Hinshaw G., Smoot G.F.,
Wright E.L., 1996, ApJ, 464, L5.



\bibitem[]{PWMAP03}
Komatsu, E. \etal 2003, preprint (astro-ph/0302223).

\bibitem[]{LS93}
Landy, S.D. \& Szalay, A.S., 1993, ApJ, 412, 64

\bibitem[]{PWMAP03}
Peiris H.V. \etal, 2003, preprint (astro-ph/0302225).

\bibitem{seljak96}{Seljak, U., Zaldarriaga, M., 1996, ApJ, 469, 437}


\bibitem[]{SWMAP03}{
Spergel D.N. \etal, 2003, preprint (astro-ph/0302209).}


\bibitem[]{szpudi03}
Szapudi, I, Prunet, S., Pogosyan, D., Szalay, A., Bond, R., 2001, ApJ 548, L11

\bibitem[]{TdOH03}{
Tegmark M., de Oliveira-Costa A., Hamilton A., 2003, preprint 
(astro-ph/0302496).}











\end{thebibliography}
\end{document}